# Rectangular augmented row-column designs generated from contractions

Hans-Peter Piepho[1]  |  Emlyn Williams[2]

[1] Biostatistics Unit, Institute of Crop Science, University of Hohenheim, Stuttgart, Germany

[2] Statistical Support Network, Australian National University, Canberra ACT, Australia

**Correspondence**

Hans-Peter Piepho, Institute of Crop Science, University of Hohenheim, 70593 Stuttgart, Germany.

Email: piepho@uni-hohenheim.de

**Funding:** none

## ABSTRACT

Row-column designs play an important role in applications where two orthogonal sources of error need to be controlled for by blocking. Field or greenhouse experiments, in which experimental units are arranged as a rectangular array of experimental units are a prominent example. In plant breeding, the amount of seed available for the treatments to be tested may be so limited that only one experimental unit per treatment can be accommodated. In such settings, augmented designs become an interesting option, where a small set of treatments, for which sufficient seed is available, are replicated across the rectangular layout so that row and column effects, as well as the error variance can be estimated. Here, we consider the use of an auxiliary design, also known as a contraction, to generate an augmented row-column design. We make use of the fact that the efficiency factors of the contraction and the associated augmented design are closely interlinked. A major advantage of this approach is that an efficient contraction can be found by computer search at much higher computational speed than is required for direct search for an efficient augmented design. Two examples are used to illustrate the proposed method.

# 1. Introduction

Field experiments laid out on a rectangular grid of plots often exhibit patterns of heterogeneity between experimental units that call for blocking along both rows and columns of the layout. Row-column designs are also relevant in other contexts where experimental units are arranged on a rectangular grid, such as greenhouse experiments or studies involving assays with 96-well plates, laid out as $12 \times 8$ arrays of wells.

Most experiments laid out according to row-column designs are fully replicated. In cases, where material for most of the treatments is limited, meaning that these treatments cannot be replicated, augmented designs (Federer, 1956, 1961) may be used, in which only a small set of treatments, referred to here as checks, are replicated, whereas the remaining treatments, referred to here as test lines, are evaluated on only one experimental unit each. Such applications are common, e.g., in early-generation plant breeding trials where seed availability for the test lines is limited, precluding a fully replicated design. The original proposal for augmented designs only involved one-way blocking (Federer 1956, 1961). Extensions to row-and-column blocking were subsequently considered in several publications, including Federer and Raghavarao (1975), Lin and Poushinsky (1983), Federer and Crossa (2005), Piepho and Williams (2016), and Vo-Thanh and Piepho (2020). With augmented row-column designs, all information on the effects of row and column blocks, as well as on the residual error variance stems from the replicated checks if the linear model used for analysis has fixed treatment effects. Hence, it is important to ensure connectivity of the design through the checks.

Recently, Bailey and Haines (2025) considered the construction of augmented designs in square arrays from smaller auxiliary designs, also known as contractions (Patterson and Williams, 1976). A key idea used by the authors is a link between the average pairwise variance of treatment differences of the augmented design and its contraction. Also focusing on designs laid out on a square array of experimental units, Williams and Piepho (2025) compared cyclic designs with computer-generated designs for the contraction, finding that the generated augmented designs agree in some settings, whereas in other settings computer-generated designs provide an edge in efficiency.

Designs laid out as square arrays have their role in practice, but in many applications the requirement of a square layout is too restrictive. For example, 96-well plates have a rectangular 8 × 12 layout. Other rectangular layouts are sometimes used for a larger number of wells, such as 24 × 16 for 384-well plates. Similarly, the layout of field experiments is seldom a square but often a rectangle. Hence, there is a need for methods of constructing rectangular augmented row-column designs providing flexibility in terms of the numbers of rows and columns, as well as the number of checks and test lines, and the proportion of plots allocated to checks. Here, we consider such methods based on the idea of using contractions as proposed in Bailey and Haines (2025). The main advantage of generating an augmented design from its contraction, as compared to direct computer-aided design generation, is the speed of construction and in particular the facility to attain good average efficiency factors very quickly. Our approach is based on the computer-based generation of a near-optimal contraction from which an augmented row-column design is obtained.

The remainder of this paper is structured as follows. In the next section, we state our main result, along with the proof, and provide two illustrative examples. In the third section we consider special cases in which a direct link between the efficiency factors of an augmented design and those of its contraction can be established. The paper ends with a brief discussion and conclusion.

## 2. The main result

Our aim is to construct an efficient augmented design in a rectangle with $v$ rows and $s$ columns ($v \geq s$). In each column of the design there will be $k$ checks and $(v-k)$ test lines, i.e. $(v-k)s$ test lines overall. The number of different checks will equal $k$, and each column of the augmented design will comprise one plot for each of the $k$ checks. Each row will have $r_h$ ($h = 1 \ldots v$) checks and $s - r_h$ test lines. Where the parameters $v$, $s$ and $k$ allow, we aim for equal $r_h$ for each row of the augmented design; otherwise they will differ by no more than one.

The construction starts with a contraction for $v$ pseudo-treatments with the $h$th pseudo-treatment replicated $r_h$ times in a $k \times s$ array and we define $r' = (r_1 \ldots r_v)$. We assume that the contraction is a binary design and look for an optimal or near-optimal row-column design. Note that here we use the term 'pseudo-treatments' for the components of the contraction, rather than the more common term 'treatments,' and that these $v$ pseudo-treatments serve as an auxiliary device to allocate checks to the plots of the augmented design; they are not identical to the $v^*$ treatments of the augmented design obtained from the contraction. For the augmented design we refer to the treatments as 'test lines' and 'checks'.

From the contraction an augmented row-column design can be constructed in a $v \times s$ array as follows:

1. Label the test lines from 1 through $(v-k)s$
2. Label the checks from $(v-k)s + 1$ through $v^* = (v-k)s + k$
3. If pseudo-treatment $l$ ($l = 1 \ldots v$) is in position $(i,j)$ of the contraction ($i = 1 \ldots k, j = 1 \ldots s$), then place check $(v-k)s + i$ in position $(l,j)$ of the augmented design
4. Fill the remaining $(v-k)s$ empty positions of the augmented design with the test lines

For the augmented design we assume the model

$$y = X\tau + Z_R\rho + Z_C\gamma + \varepsilon$$

where $X$, $Z_R$ and $Z_C$ are $n \times v^*$, $n \times v$ and $n \times s$ design matrices for the effects of treatments, rows and columns, respectively, with $n = v \times s$. The information matrix is given by

$$A = u^\delta - \tfrac{1}{s}M_R M_R' - \tfrac{1}{v}M_C M_C' + uu' \text{ (John and Williams, 1995, eq 5.3)}$$

where $u$ is a vector of treatment replications, in order, with checks at the end, i.e. $u' = (1'_{(v-k)s}, s1'_k)$, $M_R = X'Z_R$, $M_C = X'Z_C$ and $u^\delta = diag(u)$. We are interested in the canonical efficiency factors (*cefs*) of $A^* = u^{-\delta/2} A u^{-\delta/2}$ (John and Williams, 1995, eq 5.5) which we can write as

$$A^* = I_{v^*} - u^{-\delta/2} X' [Z_R Z_C] \begin{pmatrix} sI_v & 0 \\ 0 & vI_s \end{pmatrix}^{-1} \begin{bmatrix} Z_R' \\ Z_C' \end{bmatrix} X u^{-\delta/2} + \tfrac{1}{v^*} J_{v^*,v^*}$$

where $I_{v^*}$ is the identity matrix of size $v^*$ and $J_{v^*,v^*}$ is the $v^* \times v^*$ matrix of ones. The $v^* - 1$ non-trivial eigen values of the central term will the same as those of

$$\begin{pmatrix} sI_v & 0 \\ 0 & vI_s \end{pmatrix}^{-1/2} \begin{bmatrix} Z'_R \\ Z'_C \end{bmatrix} X u^{-\delta} X' [Z_R Z_C] \begin{pmatrix} sI_v & 0 \\ 0 & vI_s \end{pmatrix}^{-1/2} \quad (1)$$

along with $v^* - 2v - 1$ zero eigen values.

This expression can be manipulated with the addition/subtraction of terms related to the trivial eigen vectors, e.g. those of the form $(a1'_v, b1'_s)$ for some $(a, b)$; these correspond to trivial eigen values and so, such changes do not affect the canonical efficiency factors (*cefs*) of the augmented row-column design. Hence, we can derive the form

$$B = \begin{pmatrix} sI_v & 0 \\ 0 & vI_s \end{pmatrix}^{-1/2} \begin{bmatrix} r^\delta - \frac{1}{s}W & N_C - \frac{1}{s}r^\delta J_{v,s} \\ N'_C - \frac{1}{s}J_{s,v}r^\delta & kI_s \end{bmatrix} \begin{pmatrix} sI_v & 0 \\ 0 & vI_s \end{pmatrix}^{-1/2} \quad (2)$$

where $N_C$ is the $v \times s$ column incidence matrix for the contraction and $W = N_R N'_R$ is its $v \times v$ row concurrence matrix. The $(v-1) + (s-1)$ non-trivial eigen values of $B$ are eigen values (*cefs*) of the scaled information matrix of the augmented design, namely $A^*$. These *cefs* can be then combined with $v^* - (v-1) - (s-1) - 1$ unit eigen values of $A^*$ in order to calculate the average efficiency factor of the augmented design, namely $E_{aug}$. Hence, considering that $E_{aug}$ is the harmonic mean of the *cefs,* we are therefore interested in the trace of the inverse of $B$. We now make use of a block matrix identity (Gubner, 2024) to show that the diagonal blocks of $B^{-1}$ can be written as

$$\frac{v}{k}\left[ r^\delta - \frac{1}{s}W - \frac{1}{k}\left(N_C - \frac{1}{s}r^\delta\right)J_{v,s}\right)\left(N'_C - \frac{1}{s}J_{s,v}r^\delta\right) \right]^{-1} \quad (3)$$

and

$$\frac{v}{k}\left[ I_s - \left(N'_C - \frac{1}{s}J_{s,v}r^\delta\right)\left(r^\delta - W + J_{v,v}\right)^{-1'}\left(N_C - \frac{1}{s}r^\delta J_{v,s}\right) \right]^{-1} \quad (4)$$

Note that an extra matrix $J_{v,v}$ has been added into the middle term in (4) to make it non-singular; this does not affect the relevant eigen values. After multiplying out and ignoring unimportant terms in $J_{v,v}$ we can recognize the expression within the square brackets of (3) as the row-column information matrix of the contraction (John and Williams, 1995, eq. 5.3). Let $C_v$ be the harmonic mean of the non-trivial eigen values of the expression in square brackets of (3) and $\bar{r}$ be the arithmetic average of the $r_h$ ( $h = 1 \ldots v$) pseudo-treatment replication numbers. Then we define $\bar{C}_v = C_v/\bar{r}$. Note that for equal replication numbers, $\bar{C}_v$ becomes the average efficiency factor of the row-column contraction. In a similar manner, we define $\bar{C}_s$ as the harmonic mean of the non-trivial eigen values of the expression in square brackets of (4). Expression (4) is a little more complicated but still involves components associated with the contraction and in particular there is a type of interaction between its row and column incidence matrices. With these definitions we can then establish a formula for the average efficiency factor of the augmented row-column design, i.e.

$$E_{aug} = \frac{(v^* - 1)}{\left\{v^* - (v + s) + 1 + \frac{v}{k}\left(\frac{v-1}{\bar{C}_v} + \frac{s-1}{\bar{C}_s}\right)\right\}} \qquad (5)$$

We now consider two examples of how contractions are used to produce efficient augmented designs. The examples correspond to designs that can be used for 96- and 384-well plates. All designs considered in this paper were produced using the design generation package CycDesigN (VSNi, 2025) which constructs optimal or near-optimal designs as measured by the average efficiency factor. When constructing row-column designs, the package checks that the parameters are such that a connected design is possible. For example, we want to ensure there are non-negative degrees of freedom in the fixed effects row-column analysis of variance table of the contraction, i.e.

$$ks - 1 - (k - 1) - (s - 1) - (v - 1) \geq 0 \qquad (6)$$

**Example 1**

For a contraction with $v = 12, s = 8, k = 3$ and equal pseudo-treatment replication, i.e. $r_h = 2$ for all $h$ and hence $\bar{r} = 2$. The package gives the contraction

|     | Column |   |   |   |    |    |   |    |
|-----|--------|---|---|---|----|----|---|----|
|     | 3      | 7 | 9 | 1 | 10 | 8  | 2 | 6  |
| Row | 12     | 10| 4 | 3 | 11 | 9  | 5 | 2  |
|     | 7      | 5 | 6 | 4 | 1  | 11 | 8 | 12 |

with $\bar{C}_v = 0.5739$ and $\bar{C}_s = 0.4828$. The resulting augmented row-column design is the following $12 \times 8$ array:

|     | Column |    |    |    |    |    |    |    |
|-----|--------|----|----|----|----|----|----|----|
|     | 1      | 10 | 19 | 73 | 75 | 46 | 55 | 64 |
|     | 2      | 11 | 20 | 28 | 37 | 47 | 73 | 74 |
|     | 73     | 12 | 21 | 74 | 38 | 48 | 56 | 65 |
|     | 3      | 13 | 74 | 75 | 39 | 49 | 57 | 66 |
|     | 4      | 75 | 22 | 29 | 40 | 50 | 74 | 67 |
| Row | 5      | 14 | 75 | 30 | 41 | 51 | 58 | 73 |
|     | 75     | 73 | 23 | 31 | 42 | 52 | 59 | 68 |
|     | 6      | 15 | 24 | 32 | 43 | 73 | 75 | 69 |
|     | 7      | 16 | 73 | 33 | 44 | 74 | 60 | 70 |
|     | 8      | 74 | 25 | 34 | 73 | 53 | 61 | 71 |
|     | 9      | 17 | 26 | 35 | 74 | 75 | 62 | 72 |
|     | 74     | 18 | 27 | 36 | 45 | 54 | 63 | 75 |

This has three checks (numbered 73, 74 and 75) in each column and two in each row of the array and has average efficiency factor of $E_{aug} = 0.3881$.

For comparison, we used the augmented design option in the package to generate the full array directly, giving $E_{aug} = 0.3860$. This shows that the design generated from the contraction is very competitive, providing an edge in efficiency.

**Example 2**

For a contraction with $v = 24, s = 16, k = 5$ and pseudo-treatment replications of three or four with $\bar{r} = 3.3$, the package gives the contraction

Row

| 13 | 20 | 7  | 8  | 19 | 15 | 9  | 23 | 11 | 2  | 3  | 18 | 10 | 24 | 16 | 1  |
|----|----|----|----|----|----|----|----|----|----|----|----|----|----|----|----|
| 6  | 15 | 9  | 10 | 12 | 1  | 14 | 18 | 23 | 17 | 5  | 3  | 21 | 16 | 22 | 8  |
| 21 | 4  | 19 | 22 | 15 | 5  | 11 | 2  | 17 | 9  | 8  | 14 | 24 | 6  | 13 | 18 |
| 20 | 9  | 5  | 23 | 21 | 10 | 13 | 6  | 7  | 24 | 17 | 15 | 4  | 8  | 18 | 12 |
| 1  | 22 | 6  | 19 | 16 | 2  | 10 | 4  | 21 | 12 | 20 | 7  | 3  | 14 | 17 | 11 |

with $\bar{C}_v = 0.7749$ and $\bar{C}_s = 0.7332$. The resulting augmented rectangular design is the following $24 \times 16$ array:

Row

| 309 | 20  | 39  | 58  | 77  | 306 | 115 | 134 | 153 | 172 | 191 | 210 | 229 | 248 | 267 | 305 |
|-----|-----|-----|-----|-----|-----|-----|-----|-----|-----|-----|-----|-----|-----|-----|-----|
| 1   | 21  | 40  | 59  | 78  | 309 | 116 | 307 | 154 | 305 | 192 | 211 | 230 | 249 | 268 | 286 |
| 2   | 22  | 41  | 60  | 79  | 96  | 117 | 135 | 155 | 173 | 305 | 306 | 309 | 250 | 269 | 287 |
| 3   | 307 | 42  | 61  | 80  | 97  | 118 | 309 | 156 | 174 | 193 | 212 | 308 | 251 | 270 | 288 |
| 4   | 23  | 308 | 62  | 81  | 307 | 119 | 136 | 157 | 175 | 306 | 213 | 231 | 252 | 271 | 289 |
| 306 | 24  | 309 | 63  | 82  | 98  | 120 | 308 | 158 | 176 | 194 | 214 | 232 | 307 | 272 | 290 |
| 5   | 25  | 305 | 64  | 83  | 99  | 121 | 137 | 308 | 177 | 195 | 309 | 233 | 253 | 273 | 291 |
| 6   | 26  | 43  | 305 | 84  | 100 | 122 | 138 | 159 | 178 | 307 | 215 | 234 | 308 | 274 | 306 |
| 7   | 308 | 306 | 65  | 85  | 101 | 305 | 139 | 160 | 307 | 196 | 216 | 235 | 254 | 275 | 292 |
| 8   | 27  | 44  | 306 | 86  | 308 | 309 | 140 | 161 | 179 | 197 | 217 | 305 | 255 | 276 | 293 |
| 9   | 28  | 45  | 66  | 87  | 102 | 307 | 141 | 305 | 180 | 198 | 218 | 236 | 256 | 277 | 309 |
| 10  | 29  | 46  | 67  | 306 | 103 | 123 | 142 | 162 | 309 | 199 | 219 | 237 | 257 | 278 | 308 |
| 305 | 30  | 47  | 68  | 88  | 104 | 308 | 143 | 163 | 181 | 200 | 220 | 238 | 258 | 307 | 294 |
| 11  | 31  | 48  | 69  | 89  | 105 | 306 | 144 | 164 | 182 | 201 | 307 | 239 | 309 | 279 | 295 |
| 12  | 306 | 49  | 70  | 307 | 305 | 124 | 145 | 165 | 183 | 202 | 308 | 240 | 259 | 280 | 296 |
| 13  | 32  | 50  | 71  | 309 | 106 | 125 | 146 | 166 | 184 | 203 | 221 | 241 | 306 | 305 | 297 |
| 14  | 33  | 51  | 72  | 90  | 107 | 126 | 147 | 307 | 306 | 308 | 222 | 242 | 260 | 309 | 298 |
| 15  | 34  | 52  | 73  | 91  | 108 | 127 | 306 | 167 | 185 | 204 | 305 | 243 | 261 | 308 | 307 |
| 16  | 35  | 307 | 309 | 305 | 109 | 128 | 148 | 168 | 186 | 205 | 223 | 244 | 262 | 281 | 299 |
| 308 | 305 | 53  | 74  | 92  | 110 | 129 | 149 | 169 | 187 | 309 | 224 | 245 | 263 | 282 | 300 |
| 307 | 36  | 54  | 75  | 308 | 111 | 130 | 150 | 309 | 188 | 206 | 225 | 306 | 264 | 283 | 301 |
| 17  | 309 | 55  | 307 | 93  | 112 | 131 | 151 | 170 | 189 | 207 | 226 | 246 | 265 | 306 | 302 |
| 18  | 37  | 56  | 308 | 94  | 113 | 132 | 305 | 306 | 190 | 208 | 227 | 247 | 266 | 284 | 303 |
| 19  | 38  | 57  | 76  | 95  | 114 | 133 | 152 | 171 | 308 | 209 | 228 | 307 | 305 | 285 | 304 |

This has five checks (numbered 305, 306, 307, 308 and 309) in each column and either three or four in each row of the array and has average efficiency factor of $E_{aug} = 0.6031$.

Using the augmented design option in the package to generate the full array directly gives $E_{aug} = 0.6026$; this is after an extended period of searching. This demonstrates the advantage of using contractions to generate augmented designs, especially for larger situations.

## 3. Special cases

The expression (5) shows the general connection between a contraction and the resulting augmented design. Here we discuss some special cases which allow a simplification of (5) and hence accentuate this link:

1. The pseudo-treatments in the contraction are equally replicated, say $\bar{r}$ times, i.e. $\bar{r} = ks/v$. This corresponds to the rows of the augmented design all having $\bar{r}$ checks, which is a desirable feature. In such a case $\bar{C}_v$ will equal the average efficiency factor of the row-column contraction ($E_{con}$). In Table 1 we have listed the parameters of 21 contractions for $v > s$ where pseudo-treatments are all replicated $\bar{r}$ times. These contractions generate augmented designs where:

    (i) The number of checks is $k$ = 3, 4 or 5, and there are $k$ check plots per column of the augmented rectangular array.

    (ii) The percentage of checks in the $v \times s$ array is between 15 and 25%; this is in fact the ratio $k/v$ as a percentage.

2. Table 1 list values for $E_{con}$, $\bar{C}_s$ and the average efficiency factor for the dual design of the contraction ($E_{dual}$) when it is considered just as a column design. In 13 cases $\bar{C}_s = E_{dual}$ since $WN_C = \bar{r}^2 J_{v,s}$. Thus, the expression inside the square brackets of (4) collapses to the equivalent of that for the dual of the column design for the contraction. This is related to the property of general balance (Nelder, 1965), i.e. the row and columns concurrence matrices of the contraction commute (Williams and Bailey, 2025). The final column in Table 1 lists values for $E_{aug}$, as can be calculated from (5).

We note that when features 1. and 2. hold, the objective of optimizing $E_{con}$ and $E_{aug}$ are closely aligned. For example, when generating a row-column design, one may first maximize the average efficiency factor for just the columns and then with columns fixed, try to optimize the whole design. This strategy is particularly valuable for larger designs. With this approach, because the non-unit *cefs* of the dual column design are the same as those of the column design (John and Williams, 1995), using $E_{con}$ as a criterion for optimization directly relates to the optimization of $E_{aug}$. Furthermore, for designs with general balance, the *cefs* for the column design also comprise those for the row-column design. In general, though, using a package to search for near-optimal contractions will result in near-optimal augmented designs.

Table 1. Properties of contractions with equal replication of pseudo-treatments and average efficiency factors for the resulting augmented designs.

| k | v | s | $\bar{r}$ | $E_{con}$ | $\bar{C}_s$ | $E_{dual}$ | $E_{aug}$ |
|---|---|---|---|---|---|---|---|
| 3 | 12 | 8  | 2 | 0.5739 | 0.4828 | 0.4828 | 0.388112 |
| 3 | 15 | 10 | 2 | 0.5359 | 0.4424 | 0.4467 | 0.368217 |
| 3 | 18 | 12 | 2 | 0.5135 | 0.4176 | 0.4205 | 0.356396 |
| 4 | 16 | 8  | 2 | 0.6618 | 0.5385 | 0.5385 | 0.450683 |
| 4 | 16 | 12 | 3 | 0.7547 | 0.7097 | 0.7097 | 0.560000 |
| 4 | 18 | 9  | 2 | 0.6479 | 0.5111 | 0.5111 | 0.441030 |
| 4 | 20 | 10 | 2 | 0.6423 | 0.5000 | 0.5000 | 0.437095 |
| 4 | 20 | 15 | 3 | 0.7339 | 0.6825 | 0.6825 | 0.549752 |
| 4 | 22 | 11 | 2 | 0.6338 | 0.4851 | 0.4851 | 0.431698 |
| 4 | 24 | 12 | 2 | 0.6310 | 0.4793 | 0.4793 | 0.429763 |
| 4 | 24 | 18 | 3 | 0.7203 | 0.6652 | 0.6652 | 0.543467 |
| 4 | 26 | 13 | 2 | 0.6232 | 0.4688 | 0.4688 | 0.425538 |
| 5 | 20 | 8  | 2 | 0.6976 | 0.5453 | 0.5453 | 0.480081 |
| 5 | 20 | 12 | 3 | 0.7881 | 0.7201 | 0.7213 | 0.590627 |
| 5 | 20 | 16 | 4 | 0.8244 | 0.7993 | 0.8000 | 0.646791 |
| 5 | 25 | 10 | 2 | 0.6966 | 0.5294 | 0.5294 | 0.476348 |
| 5 | 25 | 15 | 3 | 0.7780 | 0.6992 | 0.7000 | 0.584213 |
| 5 | 25 | 20 | 4 | 0.8096 | 0.7801 | 0.7808 | 0.640252 |
| 5 | 30 | 12 | 2 | 0.6867 | 0.5038 | 0.5038 | 0.468846 |
| 5 | 30 | 18 | 3 | 0.7669 | 0.6805 | 0.6814 | 0.578506 |
| 5 | 30 | 24 | 4 | 0.7988 | 0.7661 | 0.7665 | 0.635813 |

## 4. Discussion and conclusion

This paper has proposed a method to obtain an augmented row-column design from a contraction, where the contraction is a computer-generated near optimal design for pseudo-treatments corresponding to the rows of the final design. It generalizes a method previously proposed by Bailey and Haines (2025) for square arrays to rectangular arrays. Their method is covered as a special case by our general result.

The extension to rectangular arrays greatly increases the options for the layout of the augmented design. However, there are some restrictions. For example, each column must have each check appearing on exactly one of the plots. This means that the proportion of plots allocated to checks must equal $k/v$, meaning that for a given number of rows $v$ the choice of the number of checks determines the proportion of plots devoted to checks. If one wants to set the proportion of plots allocated to checks to a specific value, then the number of rows needs to be chosen accordingly. These choices need to be made appreciating that both $k$ and $v$ must be integers, which limits the possible values their ratio, and hence the proportion of plots devoted to checks can take.

By comparison, there is relatively large flexibility regarding the number of test lines the design can accommodate. That number needs to equal the product $(v - k)s$, so for given choice of $v$ and $k$ one can adjust the number of columns $s$ such that the desired number of plots for test lines is obtained.

These restrictions suggest the following approach for choosing the dimensions of the trials, given by $(v, s, k)$:
(i) Determine the number $k$ of checks to be included in the trial.
(ii) Determine the approximate proportion of plots to be allocated to checks and find the value of $v$ yielding the ratio $k/v$ closest to the desired proportion.
(iii) Determine the number of test lines to be tested, making sure it is of the form $(v - k)s$. This determines the number of columns $s$ of the augmented design.
(iv) For the choices of $(v, s, k)$ found in (i) to (iii) generate an augmented design as described in Section 2.

To illustrate the approach, assume a breeder wants to use four checks and devote 20% of the trial to checks. Hence, the trial needs to have 20 rows. Further assume that 173 test lines are to be tested. The nearest multiple of $v - k = 20 - 4 = 16$ is 11, hence the trial needs to have 11 columns, providing space for $(v - k)s = 176$ test lines. Thus, the breeder may add three further test lines to complete the design.

Consider a second example where the layout is fixed to a 8 × 12 array on a 96-well plate. Assume that there are three checks and 25% of wells are to be allocated to checks. Hence, we choose $v = 12$. The design will accommodate $(v - k)s = (12 - 3) \times 8 = 72$ test lines.

Next, assume we use 24 × 16 = 384-well plate and want to devote about 15% of the wells to checks. Hence, we might choose $v = 24$ and $k = 4$, meaning that $k/v = 4/24 = 16.\overline{6}\%$ of the wells will be devoted to three checks. This leaves space for $(v-k)s = (24-4) \times 16 = 320$ test lines. Alternatively, if we are willing to devote a few more wells to checks, we can set $v = 16$ and $k = 3$, such that $k/v = 3/16 = 18.75\%$ and for $(v-k)s = (16-3) \times 24 = 312$ test lines can be accommodated.

If the requirements for a planned augmented design do not allow finding a suitable specification ($v$, $s$, $k$), one can always resort to direct generation of an augmented row-column design, at the cost of more computing time (Piepho and Williams, 2016; Vo-Thanh and Piepho, 2020). In future work, it may be explored if the idea to generate a design from a contraction can be extended to provide even more options, such as dropping the requirement to have each check present on one plot per column of the augmented row-column design.